\documentclass[twocolumn]{biophys}
\usepackage{helvet,times}
\usepackage{bm,textcomp}

\jno{kxl014} 
\gridframe{N}
\cropmark{N}

\doi{doi: 10.1529/biophysj.106.090944}

\usepackage{amsmath,graphicx}
\usepackage[round,numbers,sort&compress]{natbib}
\usepackage{authblk}

\begin{document}

\setcounter{page}{1} 

\title{Focal adhesion kinase -- the reversible molecular mechanosensor}

\author{S. Bell and E. M. Terentjev } 



\begin{abstract}%
{Sensors are the first element of the pathways that control the response of cells to their environment. After chemical, the next most important cue is mechanical, and protein complexes that produce or enable a chemical signal in response to a mechanical stimulus are called mechanosensors. There is a sharp distinction between sensing an external force or pressure/tension applied to the cell, and sensing the mechanical stiffness of the environment. We call the first mechanosensitivity of the 1st kind, and the latter mechanosensitivity of the 2nd kind. There are two variants of protein complexes that act as mechanosensors of the 2nd kind: producing either a one-off or a reversible action. The latent complex of TGF-$\beta$ is an example of the one-off action: on the release of active TGF-$\beta$ signal, the complex is discarded and needs to be replaced. In contrast, focal adhesion kinase (FAK) in a complex with integrin is a reversible mechanosensor, which initiates the chemical signal in its active phosphorylated conformation, but can spontaneously return to its closed folded conformation. Here we study the physical mechanism of the reversible mechanosensor of the 2nd kind, using FAK as a practical example. We find how the rates of conformation changes depend on the substrate stiffness and the pulling force applied from the cell cytoskeleton. The results compare well with the phenotype observations of cells on different substrates.  }

{Insert Received for publication Date and in final form Date.}

{emt1000@cam.ac.uk}
\end{abstract}

\maketitle 

\section{Introduction}

Cells exist within a complex and varying environment. To function effectively, cells must collect information about their external environment, and then respond appropriately. Cell environment has a profound effect on cell migration and cell fate. It is also a major factor in metastasis of certain cancers~\cite{Provenzano2009,Barcus2013}.

Sensing is the first part of the chain of events that constitute the cell response to external stimuli. Cells respond to a variety of cues; both chemical and mechanical stimuli must be transduced inside the cell. Mechanosensors are protein complexes that produce responses to mechanical inputs~\cite{Bershadsky2006,Bershadsky2009}. There are two distinct types of mechanosensing: reacting to an external force, or sensing the viscoelastic properties of the cell environment. We call the first mechanosensitivity of the 1st kind, and the latter mechanosensitivity of the 2nd kind.

Mechanosensitive ion channels (MSC), such as alamethicin \cite{msc3}, are an example of mechanosensors of the first kind. MSCs exist in all cells and provide a non-specific response to stress in a bilayer membrane \cite{msc1,msc2}. Local mechanical forces could be produced by many external factors, but MSC operation appears to be universal and quite simple. The ion channel is closed at low tension, opening as the tension exceeds a certain threshold, allowing ions to cross the membrane. Traditionally, MSC operation is understood as a two-state model. There is a balance of energy gain on expanding the `hole' under tension, and the energy penalty on increasing the hydrophobic region on the inner rim of the channel exposed to water on opening. These two-state systems (open/closed, or bonded/released) with the energy barrier between the states depending on applied force, are common in biophysics \cite{Evans1997,Bruinsma}. Rates of transition in these systems are often calculated using the `Bell formula' \cite{Bell1978}, which has them increasing exponentially with the force. This is just the classical result of Kramers and Smoluchowski \cite{Kramers,Hanggi}, but it is invalid in the limit of high forces or weak barriers.

Mechanosensitivity of the 2nd kind is different in nature. The sensor has to actively measure the response coefficient (stiffness in this case, or matrix viscosity in the case of bacterial flagellar motion). On macroscopic scales (in engineering or rheometry) we can do this with two separate measurements: of force (stress) and of position (strain), or we could contrast two separate points of force application. One could also use inertial effects, such as impact or oscillation, to measure the stiffness or elastic constant of the element. None of these options are available on a molecular scale. The single sensor complex cannot measure relative displacements in the substrate, and the overdamped dynamics prevents any role of inertia. As a result, important biophysical work on focal adhesion complexes \cite{Schwarz2006} had to resort to an idea of dynamically growing force (or force-dependent velocity) applied to the proximal side of the two-spring sensor. Other important work \cite{Storm2013} also relies on the dynamics of applied force with an elaborate construct of `catch-bonds' whose stability increases with pulling force. In reality, the cell cytoskeletal filaments exert a pulling force that is constant on the time-scales involved. Further, the internal observable needed to sense stiffness in the catch-bond model (namely, fraction of unbound integrin-ECM bonds at a focal adhesion) does not have a clear downstream measurement process associated with it.

In an earlier study \cite{Rigozzi2014}, we addressed the problem of how a mechanosensor of the 2nd kind should work, by developing a physical mechanism with a similar action to the two-spring model of Schwarz et al. \cite{Schwarz2004,Schwarz2006}. That work focussed on the latent complex of TGF-$\beta$ \cite{Khalil1999,Tomasekn2002,shi2011}, which is an irreversible one-off sensor: after the latent complex is `broken' and active TGF-$\beta$ released, the whole construct has to be replaced. Here we apply these ideas to a reversible mechanosensor: protein tyrosine-kinase, now called focal adhesion kinase (FAK) \cite{Burridge1988,Parsons1992,Tilghman2005,Eck2007}. As the name suggests, FAK is abundant in the regions of focal adhesions \cite{Parsons1992}, which are developed in the cells on more stiff substrates, often also associated with fibrosis: the development of stress fibers of bundled actin filaments connecting to these focal adhesions and delivering a substantially higher pulling force. FAK is also present in cells of soft substrates in spite of the lack of any focal adhesions, and also in the lamellipodia during cell motility \cite{Schlaepfer2005,Tilghman2005,Sheetz2}.  Phosphorylation of tyrosine residues of FAK is well known as the initial step of at least two signalling pathways of mechanosensing \cite{Provenzano2011}, leading to the cell increasing production of smooth muscle actin, and eventually fibrosis.

\section{Methods}

To achieve our aim of developing a self-consistent physical model of a molecular mechanosensor, we first need to have an extensive overview of the biological system, or the series of elements transmitting force between the cytoskeleton and the ECM. The system starts with the activated integrins binding to ECM, and follows to a group of cytoplasmic proteins that bridge between integrin and the cytoskeletal actin filaments. These in turn are assumed to provide the pulling force by the action of myosin. Among these proteins is FAK (along with talin, paxillin and vinculin), which we discuss in greater detail -- identifying the conformational transition associated with its activation, and how the effective free energy of such a protein must evolve on conformational change. 

We then proceed to the main focus of this work -- to construct the physical model that includes the viscoelastic response of the ECM and the thermally activated response of the protein mechanosensor. It turns out that both thermal activation (thermal noise) and viscous damping are essential in both elements of the mechanical chain. We outline how to `solve' this physical problem, that is, derive the effective rate of FAK opening and activation (as well as the reverse rate of its auto-inhibition) using the methods of stochastic Kramers theory. We present and test these rates in the Results section.

\subsection*{Biological system}

\begin{figure}[b]
\centering
\includegraphics[width=.9\columnwidth]{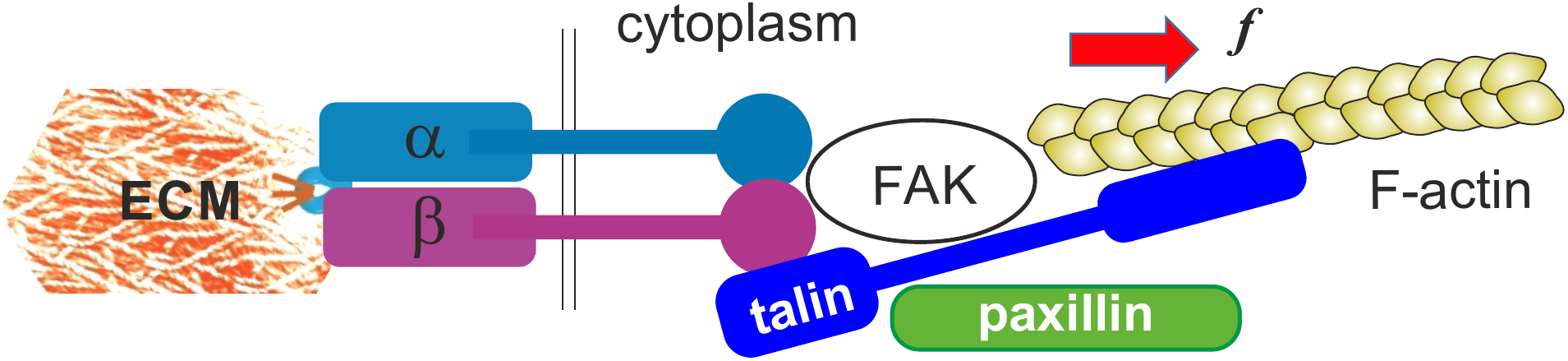}
\caption{{The chain of force transduction from the F-actin terminators of the cytoskeleton, through several associated proteins, passed on to the activated $\beta$ integrin binding to ligands of the deformable ECM.} }
\label{fig0}
\end{figure}

A sensor is a device that detects or measures a physical property and records, indicates, or otherwise responds to it. An important characteristic of any sensor is its proportional response to the input signal; in this aspect, a sensor is not a relay, which is a device that switches on/off response on receiving a sufficient level of input signal. In the case of mechanosensors of the 2nd kind, the property that we need to measure is the stiffness (elastic modulus) of the extra-cellular matrix (or other environment the cell is immersed in). To probe the modulus of a medium, a force has to be applied to it, either as a local point source, or as distributed stress. In the cell the source of this force is the actin-myosin activity of cytoskeleton. Therefore, we need to trace the series of connected devices, from the point of force origin (F-actin) to the point of its application at ECM.

Figure \ref{fig0} illustrates this force chain, which has been reproduced in a large number of important publications in this field \cite{Parsons2008,Tarone2006,Giannone2006,Eck2010,Haller2016}. As well as FAK, there are several important players that we should also consider: integrins, talin, paxillin, and the cytoskeleton. How do these components each contribute to the function of the complex?

The integrin family of transmembrane proteins link the extracellular matrix (ECM) to the intracellular actin cytoskeleton via a variety of protein-tyrosine kinases, one of which is FAK \cite{Geiger2003}. Integrins are aggregated in focal adhesions, and they mediate the cell interaction with ECM \cite{Bershadsky2006}. Activation of integrins is required for adhesion to the substrate; active integrins acquire ligand affinity and bind to the proteins of the ECM. It is well established that integrin activation and clustering leads to FAK activation and the subsequent signalling chain of mechanosensing and cytoskeletal remodelling, e.g. see the review by Parsons \cite{Parsons2003}. There is a large body of literature on integrins, with definitive reviews by Hynes \cite{Hynes1992,Hynes2002} explicitly stating that integrins are the mechanosensors. However, activated integrins possess no further catalytic activity of their own, and so can can only act in isolation as a switch, which is not the proportional response required for sensor design. A good summary by Giancotti \cite{Giancotti}, while talking about `integrin signalling', in fact, shows schemes where FAK is the nearest to cytoskeletal actin filaments. The important work by Guan et al. \cite{Guan1991,Guan1992} establishes a clear correlation chain of extracellular fibronectin -- transmembrane integrins -- intracellular FAK, but offers no reason to assume that integrin is the sensing device on this chain.

This lack of clarity regarding the specifics of integrin engagement and FAK activation arises from the lack of a detailed physical model: we simply do not understand in detail at a molecular/physical level how FAK is activated. One possibility, explored by U. Schwartz \cite{Schwarz2004,Schwarz2006}, is that clusters of activated integrins always activate FAK and generate the mechanosensing signal that leads to the increasing F-actin pulling force. As some of the integrins are broken off their ECM attachment, the associated FAK signal reduces, regulating the further force increase -- and that is the action of the focal adhesion mechanosensor complex.

In this paper we propose a different mechanism, where the activation of FAK is dependent on cytoskeletal tension and ECM stiffness, and the integrin (along with other members of the force chain in Fig. \ref{fig0}) is merely playing a role of force transducer. Of course, without the  activated integrin there would be no force transduction to ECM, and no mechanosensing. Here, we look at each individual integrin-FAK sensor, as opposed to exploring the role of clustering. This is clearly a shortcoming, as clustering is definitely an important aspect of the process: allostery of integrins (and associated FAK) must have a role in the  signalling process, as  in chemotaxis \cite{dukebray1,dukebray2}. This will have to be a topic of further study, while the present paper focuses on the physical model of individual FAK sensor operation.

There is a clear indication that phosphorylation of FAK is a key step in the mechanosensing process \cite{Bershadsky2009}.  Indeed, Schaller et al. \cite{Parsons1992} state  that FAK phosphorylation is the initial step of signalling, and show evidence that crosslinking integrins and ECM (i.e. making the `substrate' stiffer) leads to an enhanced FAK phosphorylation, while conversely, a damage to integrin is connected with a reduced activation of FAK.

\begin{figure}
\centering
\includegraphics[width=.7\columnwidth]{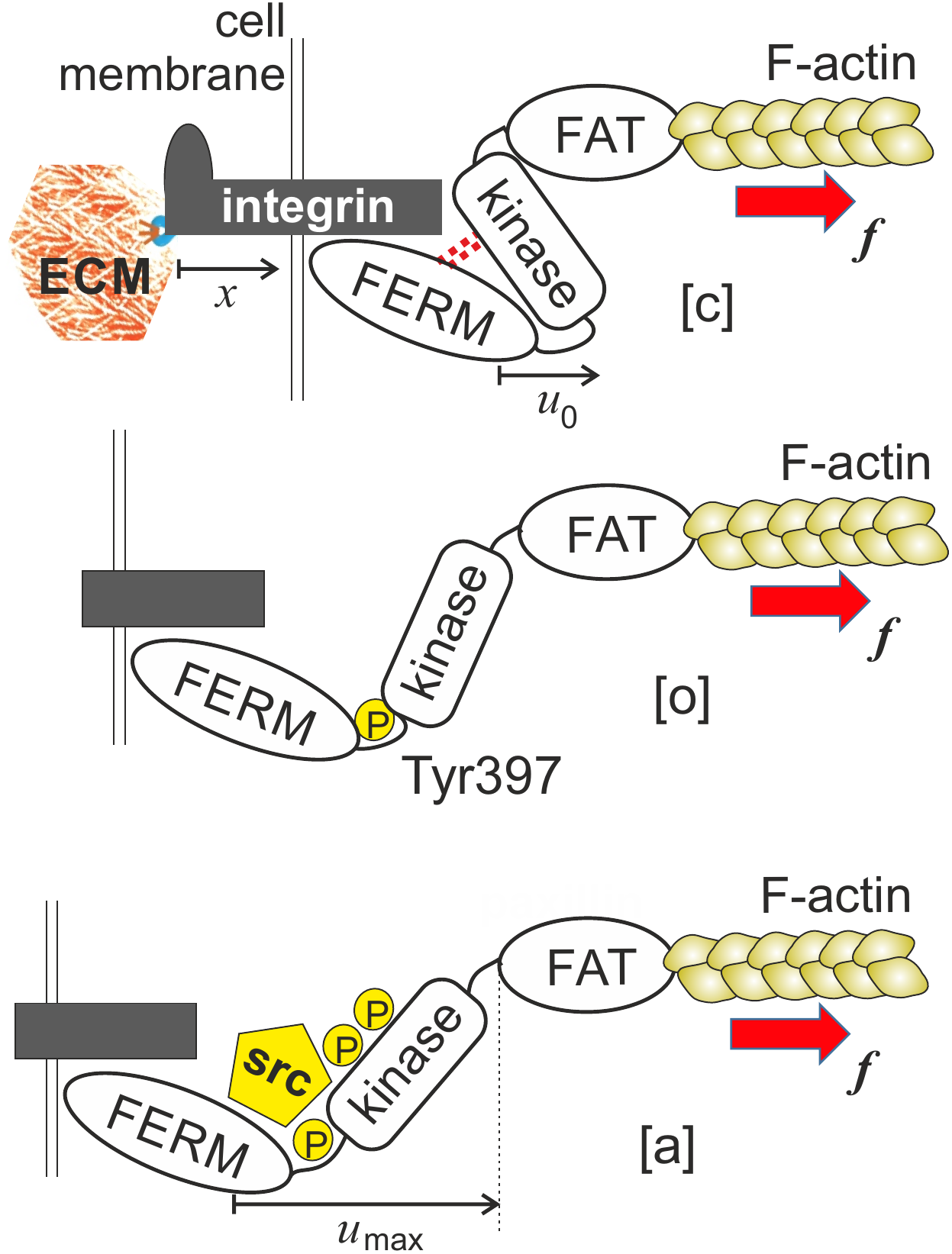}
\caption{Schematic representation of FAK conformations. The FERM domain of FAK is associated with the integrin-talin assembly, near the cell membrane. while the FAT domain is associated the actin binding site \cite{Haller2016}. The pulling force is transmitted through this chain to the FERM-kinase physical bond.  In the closed state [c] the kinase domain is inactive and the whole FAK protein is in its native low-energy state. Once the physical bond holding the FERM domain and the kinase together is broken, the protein adopts the open conformation [o]. In the open state, first the Tyr397 site spontaneously phosphorylates, which in turn allows binding of Src and further phosphorylation of the kinase - turning it into the active state [a], see \cite{Schaller2008,Guan2009,Dunty2004}.  }
\label{fig1}
\end{figure}

In the native folded state of FAK, the FERM domain (the N-terminal of the protein) is physically bonded to the catalytic domain (kinase) \cite{Eck2007,Dunty2004}; we call this closed state, [c]. A conformational change occurs, which we shall call a transition to an open state, [o], when this physical bond is disrupted and the kinase separates from the FERM domain, see Fig. \ref{fig1}. Note that because there is a peptide chain link ([362-411] segment) between the FERM and kinase domains, they remain closely associated even after the conformational change -- this is what makes FAK a reversible mechanosensor. The activation of the catalytic domain occurs in two steps: first the Tyr397 residue phosphorylates, which then allows binding of the Src kinase \cite{Schlaepfer1997}, which in turn promotes phosphorylation of several other sites of the catalytic domain (Tyr407, 576, 577, 861 and 925), making FAK fully activated. There is also a process involving p130cas, acting as a kinase substrate, involved in generating the response of activated FAK \cite{Sheetz2006,Water2007}. We shall call this state [a] in the subsequent discussion. 

 Recent work  \cite{Haller2016} explicitly confirms the critical role of tension, delivered from the actin cytoskeleton to FAK/integrin and involved in mechanosensing. A key role in this system is played by talin. There are many papers investigating the correlation of talin (as well as paxillin) with $\beta$-integrin and FAK, but recent advances clearly show that talin is capable of high stretching by a tensile force \cite{talin2011,talin2016}, implying a function similar to that of titin in muscle cells (acting as an extension-limiter). It is also now clear that the immobile domain at the N-terminal of talin is associated with integrin, as well as the FERM domain of FAK \cite{talin2011,talin2016}, while the C-terminal of talin is associated with paxillin and the  focal adhesion targeting (FAT) domain (C-terminal) of FAK. The actin filaments of the cytoskeleton exert a pulling force on this zone. Talin, therefore, acts as a scaffold for other proteins to arrange around, but more importantly, allows force to be transmitted from the cytoskeleton to the ECM, via integrins. All of these established facts are  consistent with the model of conformational change in FAK sketched in Fig. \ref{fig1}, where the integrin is the bridging element to the ECM, with the FERM domain localised near the cell membrane and N-terminal of talin. At the opposite end, the FAT domain can be pulled away by an applied force. This model is supported by the recent computational analysis \cite{fak2015} showing that the closed and the open states of FAK are reversibly reached by increasing and decreasing of pulling force.

\begin{figure} 
\centering
\includegraphics[width=.8\columnwidth]{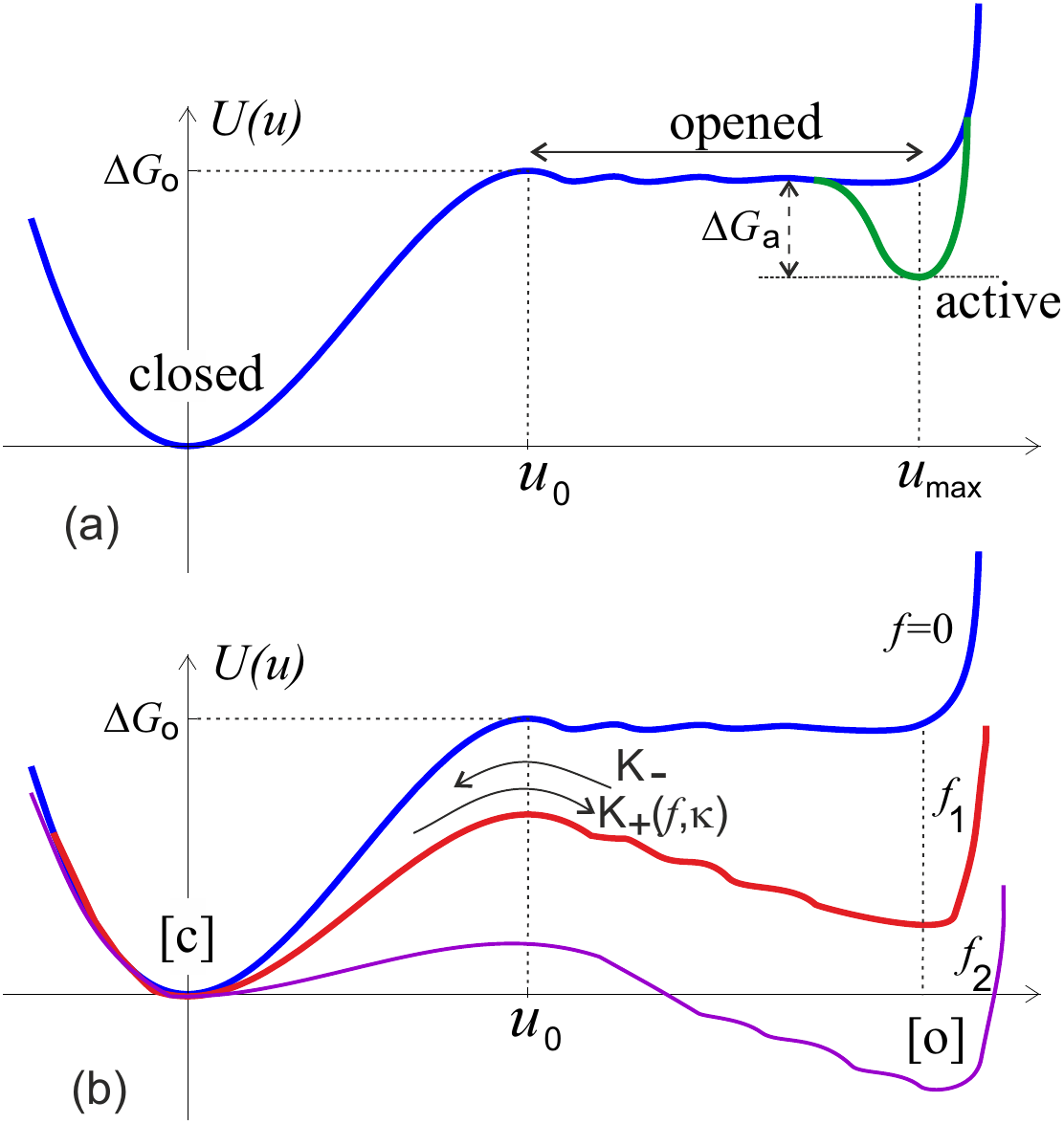}
\caption{Schematic potential energy of different FAK conformations.  (a) The force-free molecule has its native folded state [c], compare with Fig. \ref{fig1}. The binding free energy $\Delta G_\mathrm{o}$ has to be overcome to separate the kinase from the FERM domain, after which there is a range of conformations of roughly the same energy is achieved by further separating these two domains in the open state [o]. At full separation (distance $u_\mathrm{max}$) the Src binding and kinase phosphorylation lead to the active state [a] of the protein, with the free energy gain $\Delta G_\mathrm{a}$. \ (b) When a pulling force is applied to this system ($f_2 > f_1 > 0$) the potential energy profile distorts, so that both [o] and [a] states shift down in energy by the same amount of $-f\cdot u_\mathrm{max}$.}
\label{fig2}
\end{figure}

Since we shall not consider the cell motility, one has to assume that FERM domain remains fixed with respect to the ECM/integrin reference frame. That is, if there is a deformation in (soft) ECM, then this point will move accordingly, with the integrin and the local cell membrane all joined together.
In our model, to achieve the large displacement associated with the [c]$\rightarrow$[o] conformational transition of FAK, in the crowded intracellular environment, a mechanical work is expended. This mechanical energy can only come from the active cytoskeletal forces, delivered via actin filaments. 

We can now record these conformational changes in the FAK structure in the schematic plot of the `unfolding free energy', which will play the role of potential energy  $U(u)$ for the subsequent stochastic analysis of the sensor action, illustrated in Fig. \ref{fig2}. The concept of such unfolding free energy is becoming quite common \cite{Karplus2013}, when one identifies an appropriate reaction coordinate and discovers that a deep free energy minimum exists in the native folded state, with a broad range of intermediate conformations having a ragged, but essentially flat free energy profile -- before the final full unfolding rises the energy rapidly.  Figure \ref{fig2}(a) needs to be looked at together with the conformation sketches in Fig. \ref{fig1}: the native state [c] needs a substantial free energy ($\Delta G_\mathrm{o}$) to disrupt the physical bonds holding the kinase and FERM domains together. However, once this is achieved, there are only very minor free energy changes due to the small bending of the [362-411] segment \cite{Eck2007}, when the kinase and FERM domains are gradually pulled apart. This change is measured by the relative distance, which we label $u$ in the sketch and the plot. If one insists on further separation of the protein ends, past the fully open conformation [o] at $u=u_\mathrm{max}$, the protein will have to unfold at a great cost to the free energy. Binding of Src and phosphorylation (i.e. converting the [o] state into the [a] state) lowers the free energy of the fully open conformation by an amount $\Delta G_\mathrm{a}$. Note that there is no path back to the closed state, once the kinase is activated: one can only achieve `autoinhibition' \cite{Eck2007} via the [a]$\rightarrow$[o]$\rightarrow$[c] sequence. 

If we accept the basic form of the protein potential profile, as shown in Fig. \ref{fig2}(a), the effect of the pulling force $f$ applied to FAK from the actin cytoskeletal filaments is reflected by the mechanical work: $U(u) - f \cdot u$. If we take the reference point $u=0$ as the closed native conformation, then the opening barrier reduces by: $\Delta G_\mathrm{o} - f\cdot u_0$. Similarly, the free energy of the fully open state [o] becomes lower by: $\Delta G_\mathrm{o} - f\cdot u_\mathrm{max}$, see Fig. \ref{fig2}(b). Since the binding free energy of Src and phosphorylation does not depend on the applied force, the energy level of the active state [a] lowers by the same amount of  $\Delta G_\mathrm{a}$ relative to the current [o] state.

\subsection*{Stochastic two-spring model}

The two-spring model discussed in detail by Schwarz et al. \cite{Schwarz2004,Schwarz2006} and often reproduced afterwards \cite{Provenzano2011} is a correct concept, except that it needs to take into account that both the viscoelastic substrate and the sensor, described by the potential energy $U(u)$, experience independent thermal excitations. This is inevitable at the molecular level, since we are considering the mechanical damping in the substrate (as we must) and in the sensor (as we will). In the overdamped limit all forces must balance along the series of connected elements, and only thermal fluctuations -- independent in the two elements -- can create a relative displacement in the middle of this series (i.e. on the sensor). It is this relative displacement that one needs to `measure' the stiffness.

\begin{figure} 
\centering
\includegraphics[width=.75\columnwidth]{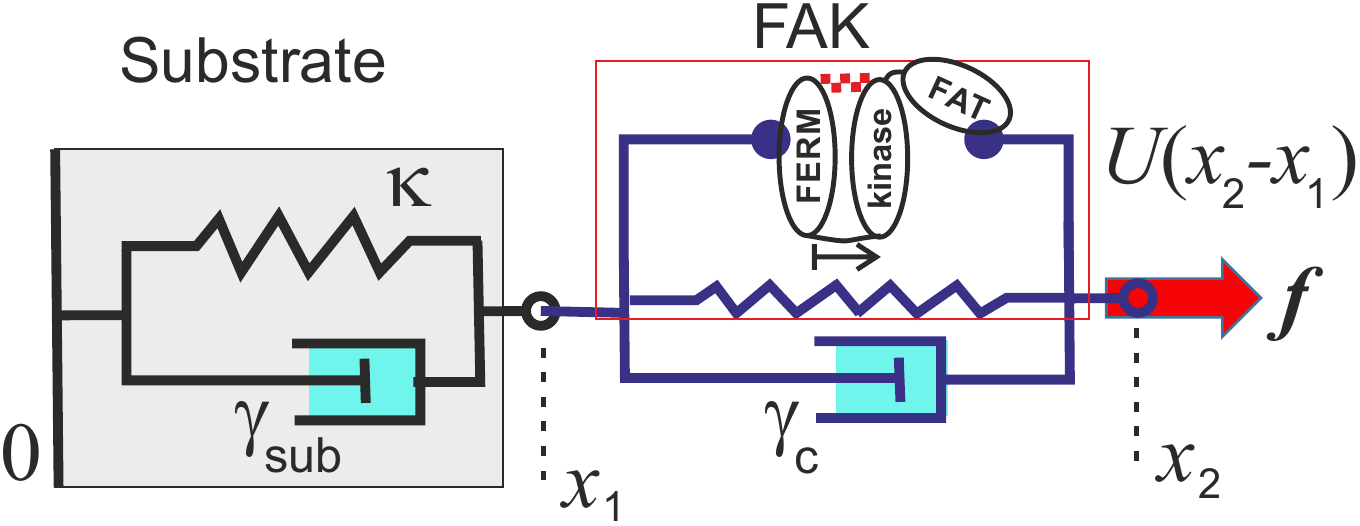}
\caption{A scheme of the 2-spring model used to produce equations \eqref{eq:lang}. The viscoelastic substrate is characterised by its elastic stiffness and stress-relaxation time given by $\gamma_\mathrm{sub}/\kappa$. The conformational change of FAK is described by a potential $U(u)$, see Fig. \ref{fig2}, and the associated relaxation time determined by the damping constant $\gamma_\mathrm{c}$.  }
\label{fig3new}
\end{figure}

Following the logic outlined in greater detail in earlier work \cite{Rigozzi2014}, we introduce two independent stochastic variables. The first is $x_1=x$, which measures the displacement of the substrate with respect to its undeformed reference state, and therefore also marks the position of the FERM domain (or the origin of the length $u$). The second is $x_2$ that measures the displacement of the far end of the kinase domain: the point of application of the pulling force $f$, see Fig. \ref{fig3new} for an illustration. These two variables satisfy  a pair of coupled overdamped Langevin equations:
\begin{eqnarray}
\gamma_\mathrm{sub} \dot{x}_1 &=&  - \kappa x_1 + \frac{\mathrm{d}U}{\mathrm{d}(x_2-x_1)}  + \sqrt{2k_BT\, \gamma_\mathrm{sub}} \cdot \zeta(t),  \nonumber \\
\gamma_\mathrm{c} \dot{x}_2 &=& - \frac{\mathrm{d}U}{\mathrm{d}(x_2-x_1)} + f + \sqrt{2k_BT\, \gamma_\mathrm{c}} \cdot \zeta(t),   \label{eq:lang}
\end{eqnarray}
where $\kappa$ is the elastic stiffness and $\gamma_\mathrm{sub}$ the damping constant of viscoelastic substrate (ECM), while $\gamma_\mathrm{c}$ is the (completely independent) damping constant for the conformational changes in FAK structure; the base stochastic process $\zeta(t)$ is assumed to be Gaussian and normalised to unity. Note that it is the difference in independent position coordinates $u=x_2-x_1$, that affects the sensor potential $U(u)$. The problem naturally reduces to a 2-dimensional Smoluchowski equation for the variables $x=x_1(t)$ for the substrate, and $u(t)$ for the FAK conformations, with the corresponding diffusion constants $D_i=k_BT /\gamma_i$, and the Cartesian components of diffusion current:
\begin{equation} \label{eq:Ji}
J_i = - \frac{k_BT}{\gamma_i} e^{-V_{\rm eff}/k_BT} \nabla_i \left(e^{V_{\rm eff}/k_B T} P \right) \;,
\end{equation}
where $P(x,u;t)$ is the probability distribution of the process, and 
\begin{equation} \label{eq:Veff}
\begin{split}
V_{\rm eff}(x_1,x_2) & = \frac{1}{2} \kappa x_1^2 - fx_2 + U(x_2-x_1)   \\
& = \frac{1}{2} \kappa x^2 - fx + U(u) - fu 
\end{split}
\end{equation}
represents the effective potential landscape over which the substrate and the mechanosensor complex move, subject to thermal excitation and the external constant force $f$.

\begin{figure}
\centering
\includegraphics[width=.8\columnwidth]{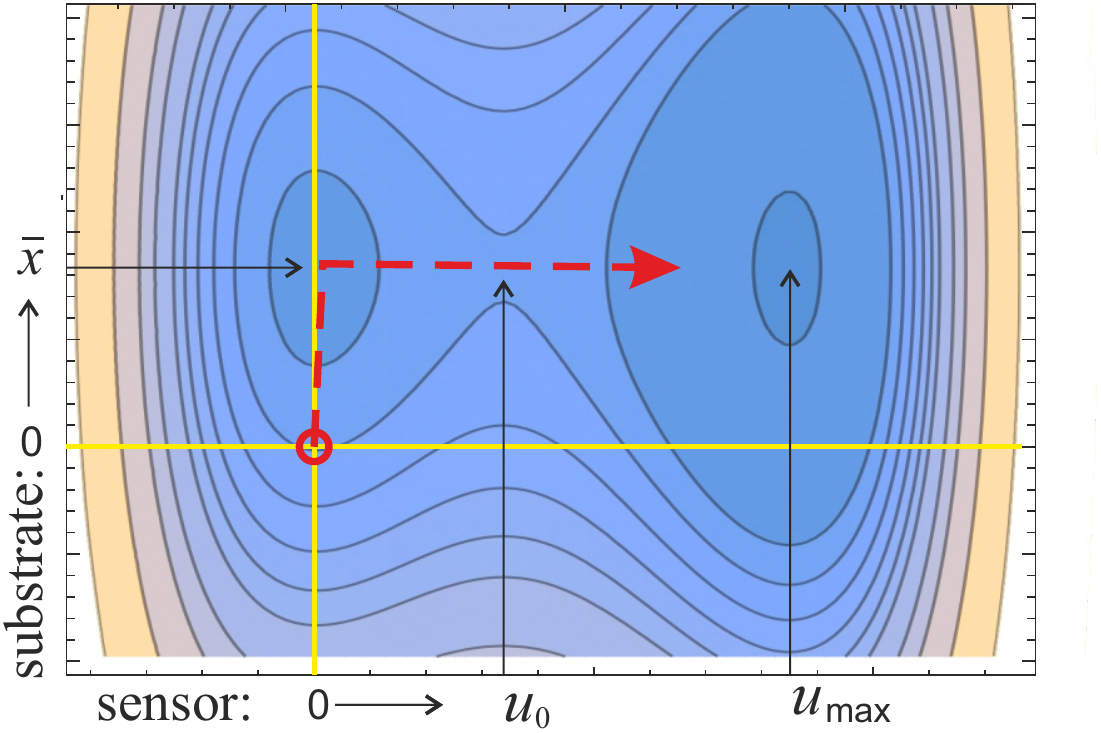}
\caption{The 2D contour plot of the effective potential $V_{\rm eff}(x,u)$ at a certain value of pulling force applied. The position of substrate anchoring has moved from $x=0$ to $\bar{x} = f/\kappa$, and the depth of the energy well of the [o] state has lowered to $\Delta G_\mathrm{o}-f u_\mathrm{max}$.  The dashed line shows the trajectory of the system evolution that leads to the opening of the [c] state. }
\label{fig4}
\end{figure}

The effective Kramers problem of escape over the barrier has been solved many times over the years \cite{Kramers, Brinkman, Hanggi, Evans1997,Dudko}. The multi-dimensional Kramers escape problem, with the potential profile not dissimilar to that in Fig. \ref{fig4} was also solved many times \cite{Langer,Dudko2}. Unlike many previous approaches, we will not allow unphysical solutions by mistreating the case of very low/vanishing barrier. In the case when the effective potential barrier is not high enough to permit the classical Kramers approach of steepest descent integrals, one of several good general methods is via Laplace transformation of the Smoluchowski equation \cite{Gardiner, Hanggi}. The compact answer for the mean time of first passage from the closed state [c] to the top of the barrier of height $Q$  a distance $\Delta u$ away is:
\begin{equation} \label{eq:tau1} 
\tau_+ = \frac{\Delta u^2}{D} \left[ \left( \frac{k_BT}{Q} \right)^2 \left( e^{Q / k_BT} -1 \right) -\frac{k_BT}{Q} \right].
\end{equation}
This is a key expression, which gives the standard Kramers thermal-activation law when the barrier is high (which is also the regime when the `Bell formula' \cite{Bell1978} is valid), but in the limit of low barrier it correctly reduces to the simple diffusion time across the distance $\Delta u$. 

\subsubsection*{Estimates of material parameters}

 In order to make plots with parameter values corresponding to a real cell, let us start with the strength of the bond holding the FERM and kinase domain in the closed (autoinhibited) state.  The  MD simulation study \cite{fak2015} estimated the energy barrier for FAK opening as $\Delta G_\mathrm{o}/k_BT \approx 25$, which is ca. $ 15$ kcal/mol at room temperature. This value appears too high, since it is known that interdomain hydrophobic interaction in such proteins is usually low-affinity \cite{bhaskara2011stability}. A reasonable value for this interdomain bonding is ca. $7$\,kcal/mol, or $\sim 11~k_BT$. We shall present the results and quantitative model predictions using this assumed magnitude of the energy barrier $\Delta G_\mathrm{o}$. 

We can also take the position of the barrier from the same study: $u_0 = 0.9$ nm, again, a reasonable value for the protein domain structure. This makes the critical force $F_c = 3\Delta G_\mathrm{o}/2u_0  \approx 70$ pN. This is a high force that is likely to unfold most proteins, and is also unlikely to be generated by a single actin filament of a cell cytoskeleton. For comparison, the force to fully unfold integrin is quoted as 165 pN \cite{BuscemiRamonetKlingberg2011}. Buscemi et al. \cite{BuscemiRamonetKlingberg2011} also quoted 40 pN as the force required to unlock the physical bond of the latent complex of TGF-$\beta$1.  Other reports investigate the force required to disrupt the fibronectin-integrin-cytoskeleton linkage, finding the value of  only  1–-2 pN \cite{Brenner,Lecuit}. For a force $f=5$ pN, the scaled non-dimensional value $\bar{f} \approx 0.1$.

We also need to estimate values of substrate stiffness. For reference, the elastic modulus of a typical collagen-rich mammalian tendon is $1.2$\,GPa \cite{Pollock1994}, of a collagen/elastic ligament: $1.1$\,Mpa \cite{Gosline2002}, and of an aorta wall: $0.8$\,MPa \cite{Bellingham2003}. Synthetic rubber has a modulus around $100$\,kPa \cite{Mithieux2004}. If a half-space occupied by an elastic medium (e.g. gel substrate or glass plate) with the Young modulus $Y$, and a point force $F$ is applied along the surface (modelling the pulling of the integrin-ECM junction, Fig. \ref{fig1}), the response coefficient (spring constant) that we have called the stiffness is given by $\kappa = (4/3)\pi Y \xi$, where $\xi$ is a short-distance cutoff: essentially the mesh size of the substrate. This is a classical relation going as far back as Lord Kelvin \cite{landauLifshitz_1986}. For a weak gel with $Y =10$ kPa, and a characteristic network mesh size $\xi =10$ nm, we obtain $\kappa = 4.2 \cdot 10^{-4}$ N/m, and the scaled non-dimensional parameter $\bar{\kappa} \approx 0.009$. 
On a stiff mineral glass with $Y = 10$ GPa, we must take the characteristic size to be a `cage' size (slightly above the size of a monomer), $\xi = 1$ nm, which gives $\kappa = 42$ N/m, and the non-dimensional parameter $\bar{\kappa} \approx 900$. A typical stiff plastic has a value about 10 times smaller. So a large spectrum of values $\bar{\kappa}$ could be explored by living cells.

Finally, we need estimates of the damping constants. The simulation study~\cite{fak2015} determined a reasonable value for the diffusion constant of the FAK complex: $D=k_{B}T/\gamma_c \approx 6 \cdot 10^{-12}\mathrm{m}^2\mathrm{s}^{-1}$. At room temperature, this gives the damping constant: $\gamma_c=7 \cdot 10^{-10}\mathrm{kg}\,\mathrm{s}^{-1}$. Then, the overall scale (`bare magnitude') of the rate $K_+$ is  $(\Delta G_\mathrm{o} / u_0^2\gamma_\mathrm{c} ) \approx 8 \cdot 10^7 \mathrm{s}^{-1}$, which means a time scale of ca. 12 ns, cf. equation \eqref{Kp2}. This `bare' time scale is compatible with available data and simulations on full and partial protein unfolding \cite{Ferscht2000}; naturally, at given bonding energy and low pulliung force the actual rate of FAK opening/activation would be much lower: the plots indicate tens of microseconds to milliseconds range.

To estimate the damping constant of the viscoelastic substrate, we assessed the characteristic time of its stress relaxation, which is the ratio $\gamma_\mathrm{sub}/\kappa$ in our parameter notation.  The order of magnitude of stress relaxation time in gels is quite long, up to $100$ s. Using the values of $\kappa$ for gels given above, the typical damping constant is calculated as: $\gamma_\mathrm{sub} \approx 0.04 \, \mathrm{kg}\,\mathrm{s}^{-1}$, and the ratio $\zeta= \gamma_\mathrm{c}/\gamma_\mathrm{sub} \sim 7 \cdot 10^{-8}$.  For stiff substrates, we need the vibration damping time in a solid glass (one must not confuse this with the creep stress relaxation, extensively studied in glasses \cite{angell} but not related to our viscoelastic response). The characteristic time we are looking for is closer to the $\beta$-relaxation time of the `cage' motion \cite{beta}, and the literature gives values in the range of $0.01$ s \cite{damping}. Combining the corresponding value of stiffness $\kappa$ discussed above gives the damping constant $\gamma_\mathrm{sub} \approx 0.4 \, \mathrm{kg}\,\mathrm{s}^{-1}$, and the ratio $\zeta = \gamma_\mathrm{c}/\gamma_\mathrm{sub} \sim 10^{-9}$ or less.

\section{Results and Discussion}

We have established a physical model for the opening of FAK under tension. Let us now apply the generic expression mean first passage time of equation \eqref{eq:tau1} to $V_{\mathrm{eff}}(x_1,x_2)$ to find the rates of this conformational change. Having established realistic parameter values, we can plot the behaviour of our model, and test its predictions against what is observed in this biological system.

\subsection*{Rate of [c]-[o] transition:  $\bm{K_+}$}

\begin{figure*}
\centering
\includegraphics[width=.8\textwidth]{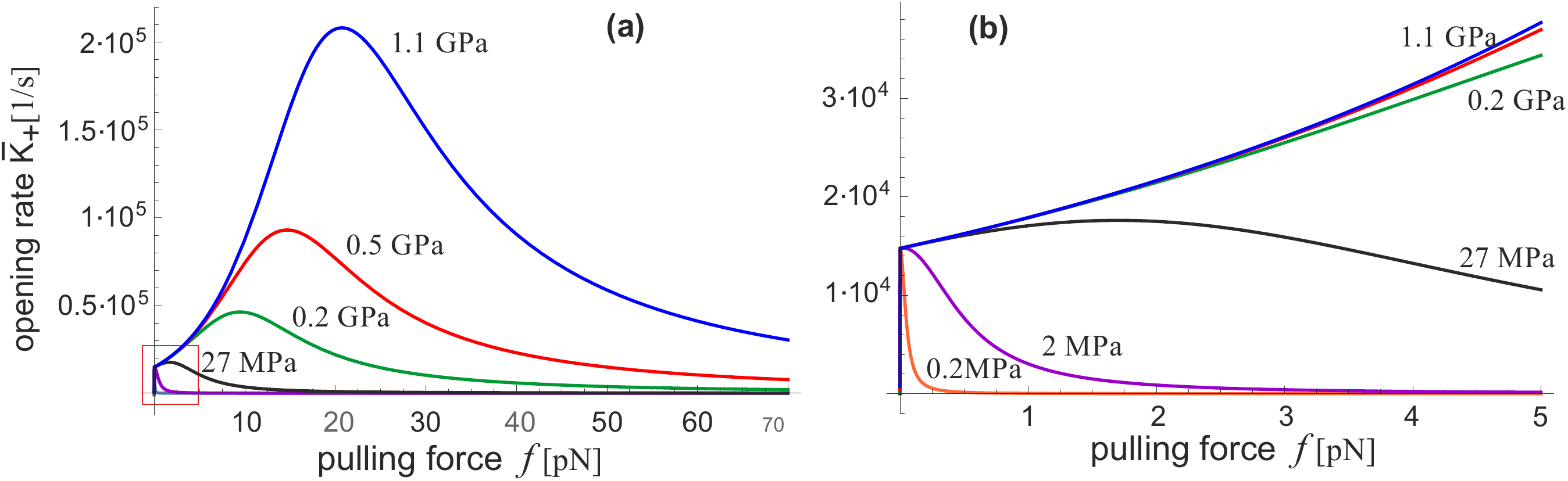}
\caption{The rate constant of FAK opening  $K_+(f,\kappa)$ is plotted as a function of the pulling force  $ f$, for several values of given  substrate stiffness labelled on the plot. Here we take the bond strength of the FERM-kinase link $\Delta G_\mathrm{o}=11 k_BT$, $u_0=0.9$ nm, and the ratio of damping constants $\gamma_\mathrm{c}/\gamma_\mathrm{sub}=10^{-7}$ (see the discussion in text about the representative values of parameters). The plot (a) illustrates the overall nature of this response, while the plot (b) zooms in the region of small forces which are biologically relevant.   }
\label{fig5}
\end{figure*}

There are many complexities regarding choosing an optimal path across the potential landscape $V_\mathrm{eff}(x,u)$, some of which are discussed in  \cite{Langer,Dudko2}, but we are aiming for the quickest way to a qualitatively meaningful answer. As such, we shall assume that the reaction path consists of two `legs': from the origin down to the minimum of the potential, which is shifted to $\bar{x} = f/\kappa$ due to the substrate deformation, and from this minimum over the saddle (barrier) into the open state of FAK conformation. The average time along the first leg is given by the equation \eqref{eq:tau1} with the distance $\Delta u=\bar{x}$ and the negative energy level $E=-f^2/2\kappa$, with the diffusion constant determined by the damping constant of the substrate:
\begin{equation} \label{tK}
\tau_\mathrm{sub} = \frac{2\gamma_\mathrm{sub}}{\kappa} +\frac{4 \gamma_\mathrm{sub} k_B T}{f^2}\left( e^{-f^2/2 \kappa k_BT} -1 \right).
\end{equation}
Here the ratio $\gamma_\mathrm{sub}/\kappa$ is the characteristic stress-relaxation time of the viscoelastic substrate \cite{Chaudhuri}, which will play a significant role in our results. Naturally, $\tau_\mathrm{sub} =0$ when there is no pulling force and the minimum is at $(0,0)$.

In the region between the minimum of $V_\mathrm{eff}$ and the potential barrier, a number of earlier papers \cite{Rigozzi2014,Dudko, Dudko2} have used the effective cubic potential to model this portion of $U(u)$. In this case, when the pulling force is applied, the barrier height is reducing as: $E = \Delta G_\mathrm{o} \left( 1-2fu_0/3 \Delta G_\mathrm{o} \right)^{3/2}$, while the distance between the minimum [c] and the maximum at the top of the barrier is reducing as:  $\Delta u =u_0  \left( 1-2fu_0/3 \Delta G_\mathrm{o} \right)^{1/2}$. Substituting these values into equation \eqref{eq:tau1}, we find the mean passage time over the barrier:
\begin{eqnarray} \label{tG}
&&\tau_\mathrm{esc} = -\frac{\gamma_\mathrm{c} u_0^2}{ \Delta G_\mathrm{o} \left( 1-\frac{2fu_0}{3 \Delta G_\mathrm{o}} \right)^{1/2}} \\
&& + \frac{\gamma_\mathrm{c} k_B T u_0^2}{ \Delta G_\mathrm{o}^2 \left( 1- \frac{2fu_0}{3 \Delta G_\mathrm{o}} \right)^2}\left( e^{\Delta G_\mathrm{o}  \left( 1- \frac{2fu_0}{3 \Delta G_\mathrm{o}} \right)^{3/2}/k_BT} -1 \right). \nonumber
\end{eqnarray}
In the limit of high barrier $\Delta G_\mathrm{o} \gg k_BT$ and small force this expression becomes proportional to $e^{- (\Delta G_\mathrm{o}-F u_0)/k_BT}$, i.e. recovers the `Bell formula' that people use widely. When the force increases towards the limit $F_c=3 \Delta G_\mathrm{o} /2u_0$, this time $\tau_\mathrm{esc}$ reduces to zero: there is no barrier left to overcome, and the minimum of $V_\mathrm{eff}$ shifts to coincide with the entrance to the [o] state. 

\begin{figure*}
\centering
\includegraphics[width=.8\textwidth]{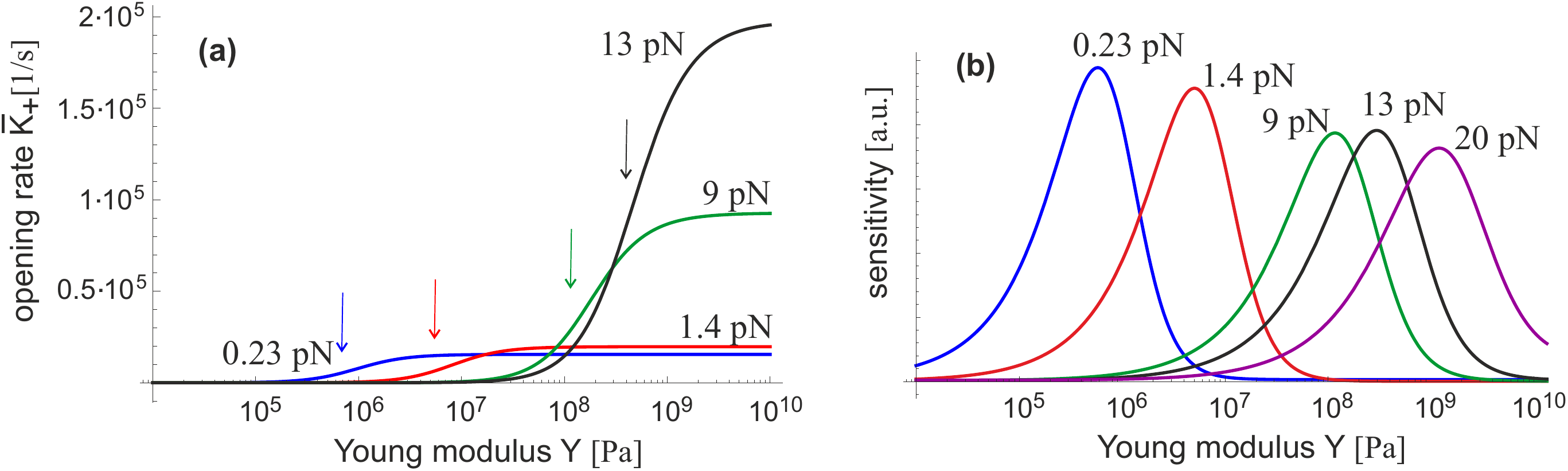}
\caption{(a) The rate constant of the [c]$\rightarrow$[o] transition ${K}_+(f,\kappa)$ plotted against the substrate stiffness (on logarithmic scale), for values of the pulling force ${f}$ corresponding to the position of the peaks in Fig. \ref{fig5}(a). As in Fig. \ref{fig5}, we take  $\gamma_\mathrm{c}/\gamma_\mathrm{sub}=10^{-7}$ for illustration. The arrows point at the inflection point on each curve, i.e. the region of maximum sensitivity.  \ (b) The plot of `sensitivity' $dK_+/ d\kappa$ for the same parameters, illustrating the maximum sensitivity range at each level of pulling force. Note that the peak in sensitivity corresponds roughly with the corresponding stiffnesses used to generate Fig. \ref{fig5}, indicating that the sensor is adaptable.}
\label{fig6}
\end{figure*}

The overall rate constant of `escape'  $K_+$ (the transition [c]$\rightarrow$[o]) is then determined as the inverse of the total time: $K_+ = (\tau_\mathrm{sub} + \tau_\mathrm{esc} )^{-1}$.  From examining equations \eqref{tK} and \eqref{tG} it is evident that the rate of FAK opening is a strong function of the pulling force $f$, but more importantly: it changes dramatically with the substrate stiffness $\kappa$. The important  exponential factor $e^{f^2/2\kappa k_BT}$  appears in $\tau_\mathrm{sub}$; it was discussed at length in \cite{Rigozzi2014} where it has emerged in a very different approach to solving a similar problem, and interpreted as an effective `enzyme effect' of the system being confined at the bottom of the potential well before jumping over the barrier.

In order to analyse and plot it, we need to scale the rate constant $K_+$ to convert it into non-dimensional values. First, we can identify a characteristic time scale of the FAK conformational change: $u_0^2\gamma_\mathrm{c}/\Delta G_\mathrm{o}$.   The two control parameters defining the opening rate $K_+$ are also made non-dimensional: scaling the force by the natural value of the FERM-kinase holding potential, $\Delta G_\mathrm{o}/u_0$, and scaling the substrate stiffness by $\Delta G_\mathrm{o}/u_0^2$. After these transformations, and some algebra, we obtain:
\begin{eqnarray}\label{Kp2}
K_+ = \left(  \frac{\Delta G_\mathrm{o}}{u_0^2\gamma_\mathrm{c}} \right)  
\frac{g \bar{f}^2 \left( 1-2\bar{f}/3 \right)^2 \zeta}{ 4  \left( 1-2\bar{f}/3 \right)^2 \Psi_1[f] + \bar{f}^2 \zeta \Psi_2[f]},
\end{eqnarray}
with shorthand notations
\[ \Psi_1[f] = \exp[-g \bar{f}^2/2 \bar{\kappa} ]  +g \bar{f}^2/2 \bar{\kappa}-1 ,  \]
\[ \Psi_2[f] = \exp[g  \left( 1-2\bar{f}/3 \right)^{3/2} ]  -g  \left( 1-2\bar{f}/3 \right)^{3/2}  -1 ,  \]
where the non-dimensional abbreviations stand for:  the energy barrier $g = \Delta G_\mathrm{o}/k_BT$, the force $ \bar{f} = f \cdot u_0 / \Delta G_\mathrm{o}$, the substrate stiffness $  \bar{\kappa} = \kappa \cdot u_0^2/\Delta G_\mathrm{o}$, and the ratio of damping constants $\zeta=\gamma_\mathrm{c}/\gamma_\mathrm{sub}$. There are several key effects predicted by this expression for the rate of FAK opening under force, while attached to a viscoelastic base, which we can examine by plotting it. 

 The effectiveness of FAK as a mechanosensor of the 1st kind, i.e. responding to an increase of applied force with a conformation change, is illustrated in Fig. \ref{fig5}. Figure \ref{fig5}(a) highlights the rapid increase in the rate that FAK opens (and its subsequent phosphorylation) on stiffer substrates. For the complex to actively probe the substrate stiffness (mechanosensitivity of the second kind), we posit that the cell remodels itself in response to FAK activation, increasing the pulling force. This increases the level of FAK activation until a maximum rate is reached. Any increase in force beyond this point decreases the rate of FAK opening. This would act as a mechanism for negative feedback, which settles the cell tension in homeostasis. The stiffer the substrate, the higher the rate of FAK activation and, accordingly, the more $\alpha$-SMA stress fibers one would find in this adjusted cell  (leading to morphological changes such as fibroblast-myofibroblast transition, or the fibrosis of smooth muscle cells). The plot \ref{fig5}(b) zooms in to the region of small forces and highlights the effect of soft substrates. On substrates with sufficiently small $\kappa$ there is no positive force that gives a maximum in the opening rate. Thus, any pulling force on the FAK-integrin-ECM chain has the effect of lowering the activation of FAK relative to the untensioned state, and so the cell does not develop any great tension in the cytoskeleton. This is consistent with the observation that cells do not develop focal adhesions on soft gels. 

Figure \ref{fig6} presents the same rate of FAK opening, but focuses on the effect of substrate stiffness. As we have shown, the possible range of parameter $\kappa$ is large, and so we plot the axis of stiffness in logarithmic units. The rate of FAK activation has a characteristic (generic) form of any sensor in that it undergoes a continuous change between the `off' and `on' states. The latter is a state of high rate of FAK opening and the subsequent phosphorylation, initiating the signaling chains leading to more actin production and increase of stress fibers. {For each cell, characterised by a specific level of pulling force, the substrate could be `too soft', meaning that FAK does not activate at all -- and also `too stiff', where the rate of activation reaches a plateau and no longer responds to further stiffening.} Between these two limits, there is a range of maximum sensitivity where the rate of activation directly reflects the change of substrate stiffness. Figure \ref{fig6}(b) highlights this by presenting the `sensitivity' directly as the value of the derivative $dK_+/d\kappa$. We see that cells with a higher pulling force (i.e. with high actin-myosin activity and developed stress fibers) are sensitive to the substrates in the stiff range. In contrast, cells that exert a low pulling force (i.e. no stress fibers, low actin-myosin activity) are mostly sensitive to soft substrates.  This is in good agreement with broad observations about the cell mechanosensitivity of the 2nd kind, and their response to substrate stiffness.  

\subsection*{Stress relaxation in substrate regulates $\bm{K_+}$}

There are many indications in the literature that not only the substrate stiffness, but also the degree of viscoelasticity (often measured by the characteristic time of stress relaxation) have an effect on cell mechanosensitivity \cite{Chaudhuri}. It is actually irrelevant what particular viscoelastic model one should use for the substrate, and certainly impossible to have a universal model covering the highly diverse viscoelasticity of gels, filament networks, and disordered solids like plastic and glass. In the spirit of our ultimately simplified viscoelastic model expressed in equations \eqref{eq:lang}, the single parameter characterising viscoelasticity could be the characteristic time scale $\gamma_\mathrm{sub}/\kappa$: this could be a measure of the actual stress relaxation time of different substrates. This would be a very short timescale in stiff solids, while complex disordered filament networks like a typical ECM would have this time measured in minutes or hours. We now find that the rate of FAK opening is strongly affected by the viscoelastic relaxation properties of the substrate.

\begin{figure} 
\centering
\includegraphics[width=.42\textwidth]{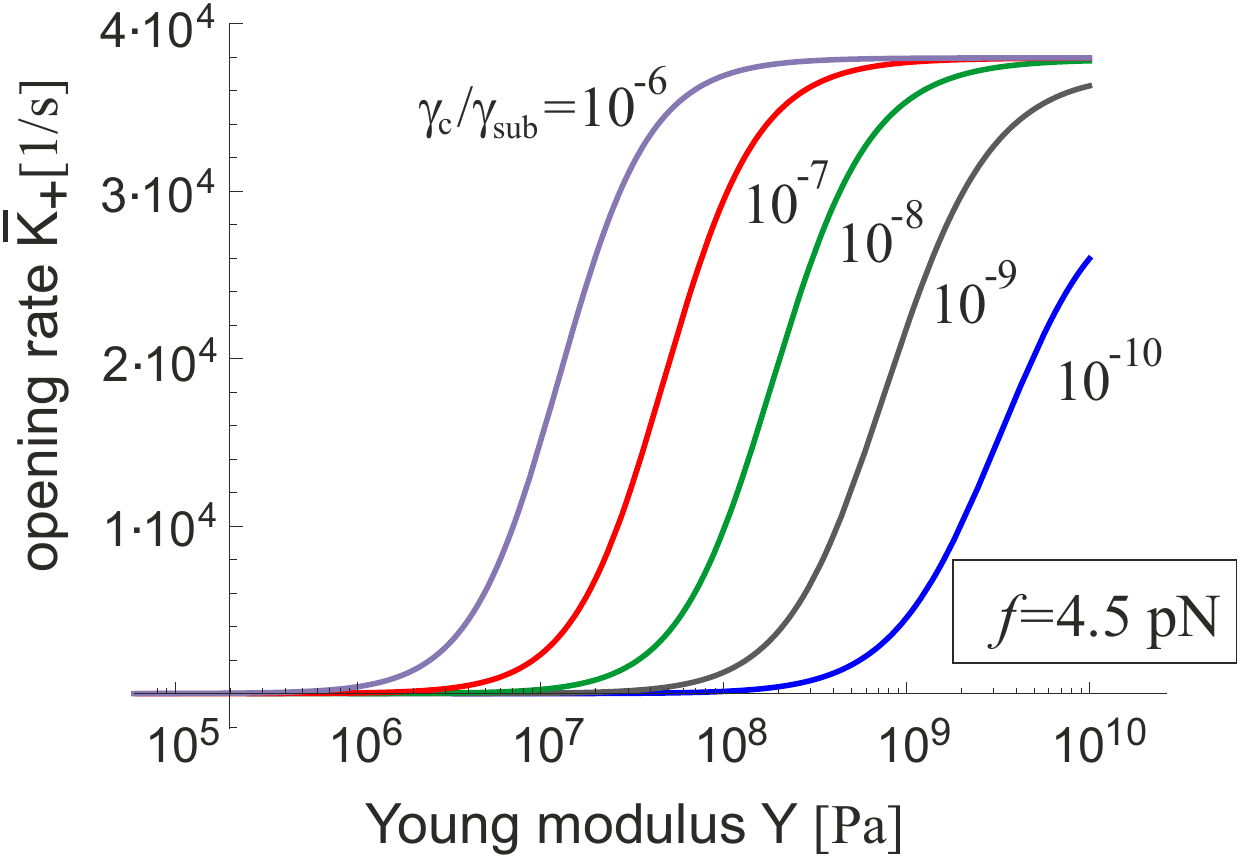}
\caption{The rate of the [c]$\rightarrow$[o] transition ${K}_+(f,\kappa)$  plotted against the substrate stiffness  for a fixed (low) value of pulling force  and a set of changing stress-relaxation properties of the substrate measured by the ratio $\gamma_\mathrm{c}/\gamma_\mathrm{sub}$, cf. equations \eqref{eq:lang} and \eqref{Kp2}.  The range of maximum sensitivity shifts to the effectively stiffer substrate range for materials with higher damping constant $\gamma_\mathrm{sub}$.  }
\label{fig7}
\end{figure}

We assume that the damping constant  $\gamma_\mathrm{c}$ of the FAK complex remains the same. In that case, Fig. \ref{fig7} shows how changing the damping constant of the substrate $\gamma_\mathrm{sub}$ (or the associated loss modulus of the viscoelastic material) can regulate the FAK mechanosensor. All curves retain exactly the same topology and amplitude, but the range of sensitivity shifts in either direction. The red curve for $\zeta=\gamma_\mathrm{c}/\gamma_\mathrm{sub} = 10^{-7}$ is the same as the red curve for ${f}=4.5$ pN in both plots in Fig. \ref{fig6}. We find that for substrates with greater stress relaxation (i.e. greater loss modulus, or $\gamma_\mathrm{sub}$, leading to the ratio $\zeta$ becoming smaller), the FAK sensor will activate at higher stiffnesses. In other words, in stiffer substrates, stress relaxation suppresses the response of a sensor with respect to a strictly elastic substrate.

\begin{figure} 
\centering
\includegraphics[width=.42\textwidth]{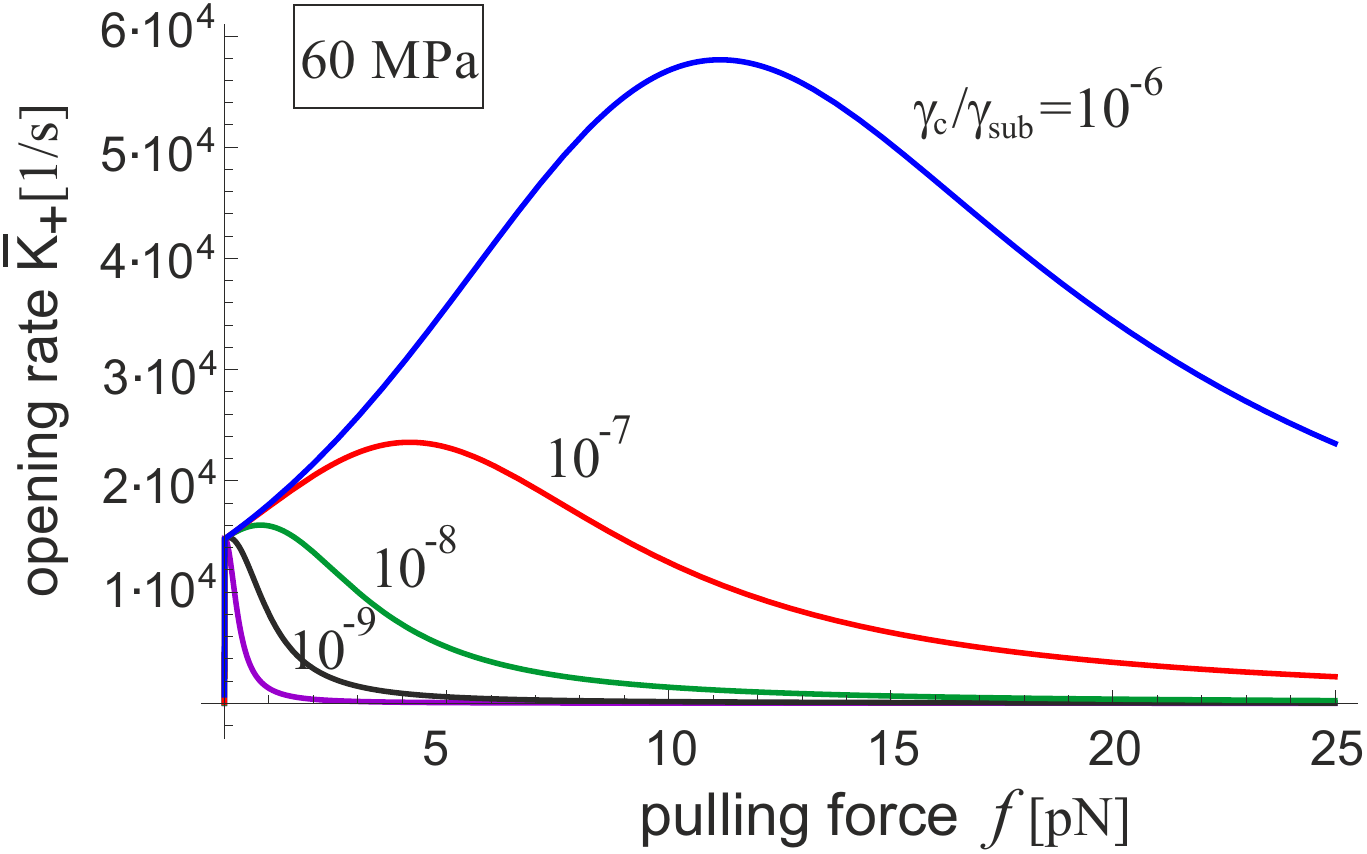}
\caption{The rate of the [c]$\rightarrow$[o] transition ${K}_+(f,\kappa)$  plotted against the pulling force ${f}$ (in the biologically relevant range of small forces) for a set of values $\gamma_\mathrm{c}/\gamma_\mathrm{sub}$ labelled on the plot representing the change in stress-relaxation characteristics of the substrate. The Young modulus of the substrate is $Y\approx 60$ MPa.}
\label{fig8}
\end{figure}

{We also see the strong effect of substrate viscoelasticity on the absolute value of rate of FAK activation $K_+$.  Figure \ref{fig8} shows a tough rubber with the Young modulus of $\sim 60$ MPa (not a completely rigid glass). A range of  $\gamma_\mathrm{sub}$ is tested, and here we see how the material with a higher loss factor (i.e. lower ratio $\zeta$) has a reduced response at a lower fulling force. This is essentially analogous to the substrate appearing `softer'. This might appear counter to the conclusion one draws from Fig. \ref{fig7} (where the range of sensitivity shifted to stiffer substrate), but one must remember that we are exploring different aspects of the same expression $K_+(f,\kappa)$: the information conveyed by Fig. \ref{fig5} is exactly the same as that in Fig. \ref{fig6}(a). }

\subsection*{Rate of [o]-[c] transition:  $\bm{K_-}$}

The free energy profile of the conformation change leading to the [o]$\rightarrow$[c] transition (i.e. the spontaneous return of FAK to its native folded conformation: the autoinhibition) is essentially described by the linear potential, see Fig. \ref{fig2}. From the reference point of [o] state, the energy barrier is $E = f(u_\mathrm{max}-u_0)$, and we should assume that the physical distance the FERM domain needs to travel remains constant: it is determined by the extent of the protein structure \cite{Eck2007,Schaller2008}. This process also does not depend on the substrate stiffness. As a result, the rate of the folding transition is the inverse of the mean first-passage time \eqref{eq:tau1} with these parameters:
\begin{equation}\label{Km}
K_-(f) = \frac{f}{\gamma_\mathrm{c}\Delta u} \left( \frac{k_BT}{f\Delta u}\left[ e^{f \Delta u/k_BT}-1 \right] - 1\right)^{-1},
\end{equation}
with the shorthand notation $\Delta u=(u_\mathrm{max}-u_0)$.  When the force is high, and the [o] state has a deep free-energy minimum generated by this external mechanical work, see Fig. \ref{fig2}(b), this rate reduces exponentially with pulling force: $K_- \approx (f^2/\gamma_\mathrm{c}k_BT)  e^{-f \Delta u/k_BT}$.  This reflects the increasing stability of the [o] state when FAK is pulled with a high force, even before it phosphorylates and further stabilises in the active state [a].  On the other hand, at vanishing force: $f \rightarrow 0$, this rate becomes $K_- \approx 2k_BT/\gamma_\mathrm{c} \Delta u^2$, which is the free-diffusion time over the distance $(u_\mathrm{max}-u_0)$, or the natural time of re-folding of the force-free open state.

We must mention several factors that would make the process of auto-inhibition more complicated, and its rate $K_-$ deviate from the simple expression \eqref{Km}. First of all, the [o] state will in most cases be quickly phosphorylated, which means there will be an additional binding energy $\Delta G_\mathrm{a}$ stabilising this conformation -- making the effective rate of autoinhibition much lower. On the other hand, there is an effect of extension-elasticity of talin \cite{talin2011,talin2016} that would provide an additional returning force acting on the FAK complex: this would make the low/zero force case fold back faster, at a higher rate $K_-$. While these are interesting and important questions that need to be investigated, at the moment we will focus on the simplest approximation to understand the universal qualitative features of FAK sensor dynamics. 

In order to be able to compare different expressions, and plot different versions of transition rates, we must identify the non-dimensional scaling of  $K_-$. Factoring the same natural time scale as we used for  $K_+$, the expression takes the form:
\begin{equation}\label{Km2}
K_- = \left(  \frac{\Delta G_\mathrm{o}}{u_0^2\gamma_\mathrm{c}} \right) 
\frac{g \bar{f}^2 }{\left( e^{g \bar{f}  \lambda} -1 \right) - g \bar{f}  \lambda},
\end{equation}
where, as before: the force is scaled as $f = \bar{f} \Delta G_\mathrm{o}/u_0$, the opening energy barrier $g = \Delta G_\mathrm{o}/k_BT$, and the ratio of two length scales (in [c] and [o] states) is labelled by the parameter $\lambda = (u_\mathrm{max}-u_0)/u_0$ (see Fig. \ref{fig1}).  We don't have direct structural information about the physical extent of FAK opening. However, taking the structural data on the separate FAK domains from the work of Eck, Schaller and Guan \cite{Eck2007,Schaller2008,Guan2009}, we make an estimate that $u_\mathrm{max} \approx 6.5$ nm, essentially determined by the double of the size of folded kinase domain, cf. Fig. \ref{fig1}. This gives $\lambda \approx 6$ and lets us plot the comparison of the two transition rates, $\bar{K}_+$ and $\bar{K}_-$. 

\begin{figure} 
\centering
\includegraphics[width=.4\textwidth]{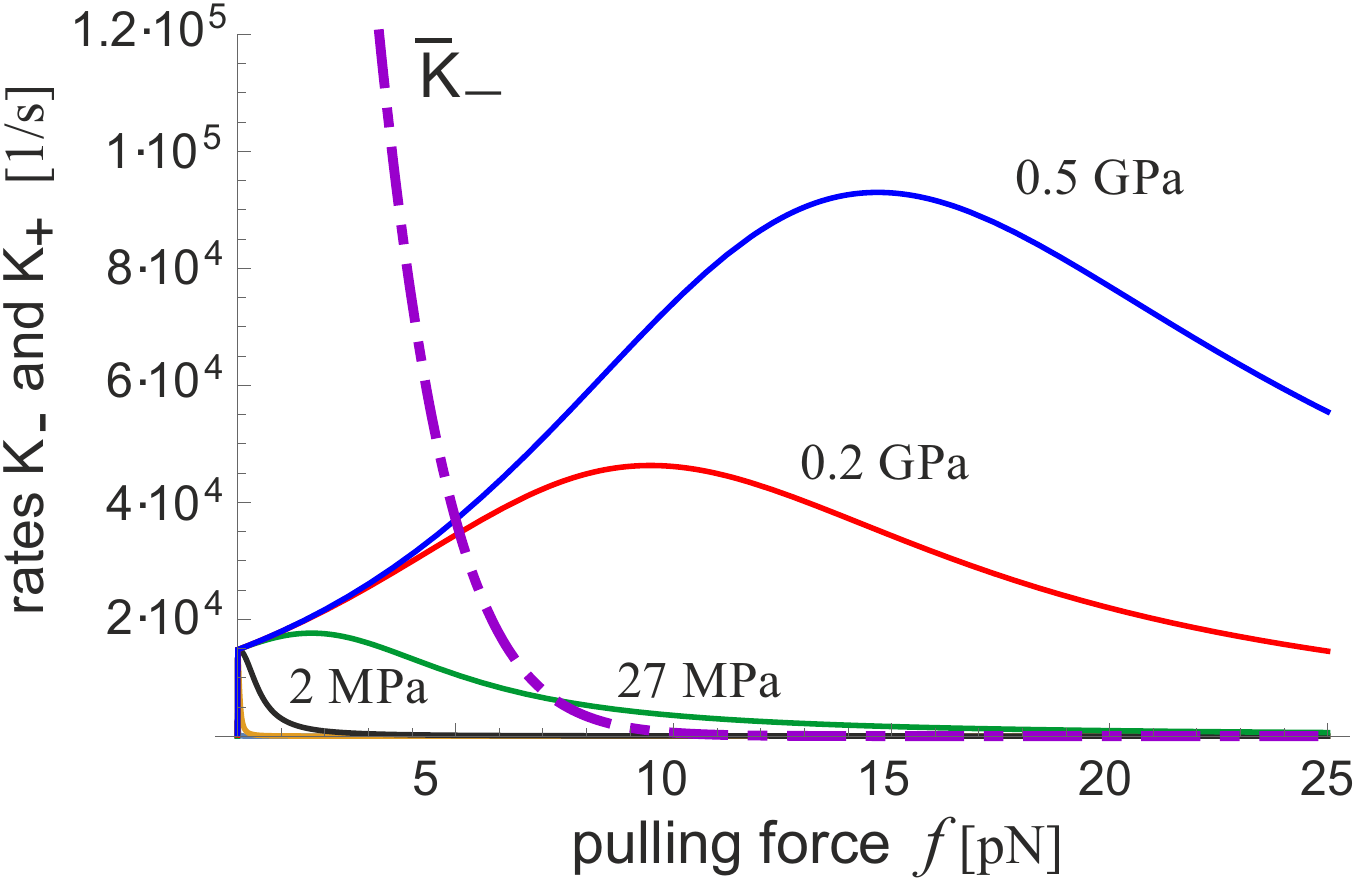}
\caption{Comparison of the opening and closing rates, ${K}_+$ and ${K}_-$, for several different substrate stiffnesses.  As in several previous plots,  $\Delta G_\mathrm{o}=11 k_BT$, $u_0=0.9$ nm, and the damping constant  ratio $\zeta = 10^{-7}$. When the cytoskeleton pulling force is too low, the rate of autoinhibition rapidly increases and one does not expect strong phosphorylation and positive feedback of mechanosensor.  }
\label{fig9}
\end{figure} 

Figure \ref{fig9} gives the transition rates, ${K}_+(f,\kappa)$ and $\bar{K}_-(f)$, plotted as a function of increasing pulling force. The rate of closing, $\bar{K}_-$, does not depend on the substrate parameters and is rapidly increasing when the [c]-[o] range of protein potential energy is flat, cf. Fig. \ref{fig2}. In this range of parameters, the product $g \bar{f} \lambda$ in the equation \eqref{Km2} is large, and the expression decays exponentially: $\exp [-g \bar{f} \lambda]$. This implies that the transition from the strongly autoinhibited population of FAK sensors to the largely activated sensors is rather sharp.
We find that the crossover force at which $K_+ \approx K_-$ is a relatively universal prediction, giving an estimate for the order of magnitude force required to keep the FAK conformation open as $f^* \approx 5$ pN.

One might be tempted, in the traditional way, to interpret the ratio of the `on' and `off' rates $K_+/K_-$ as an equilibrium concentration of closed and open/activated states. However, we must remember that this process of mechanosensing is inherently non-equilibrium, even though it might be steady-state on the time scale of sensor response. Even in the regime of very low pulling forces, when $K_- \gg K_+$, the few FAK molecules that are spontaneously open would provide the required (low) level of signal to the cell pathways. It is simply an indication of sensor reversibility: Fig. \ref{fig9} predicts that as soon as the force reduces below $f^*$, most of the FAK molecules would fold back and autoinhibit their action.

\section{Conclusions}

Figures~\ref{fig5} and~\ref{fig6} each contain lots of information, but when we link the two we uncover the true nature of this reversible mechanosensor. If we place a cell on a substrate of given stiffness $\kappa_f$ (or Young modulus $Y_f$), then according to our model, the mechanosensors will generate a positive feedback loop: increasing the rate of FAK activation, which leads (via the Erk or Rho GTPase pathways \cite{Provenzano2011}) to the increased production of actin. Assembling more F-actin, the cell will increase tension in the cytoskeleton, which will further increase the rate of FAK activation on stiffer substrates. This positive feedback goes until approximately the peak position of the curves in Fig. \ref{fig5}a, after which the further increase of tension shuts down the FAK activation response. The resulting negative feedback loop returns the cell to its homeostatic level of the cytoskeleton tension $f(\kappa_f)$, corresponding to the given substrate stiffness. Importantly, on very soft substrates (gels or soft tissues), the FAK signalling feedback is always negative and FAK autoinhibition on its own would lead to a very low cytoskeletal tension -- no focal adhesions or stress fibers are formed on such substrates. It is likely that other mechanosensors become more relevant on very soft substrates  (and in planktonic suspension), such as the TGF$\beta$ latent complex \cite{Rigozzi2014}: after all, the very name of FAK suggests its relation to focal adhesions, which only occur on stiff substrates.
This idea corresponds very well with experimental work showing that cells on sufficiently soft substrates do not form stable focal adhesions~\cite{Trappmann2012}.

Returning back to the homeostatic tension, if we look at the sensitivity of the FAK complex at this fixed level of force, $f(\kappa_f)$, we find the maximum sensitivity (identified with the peaks in Fig.~\ref{fig6}b) is very close to the actual substrate stiffness, $\kappa_f$. So, this physical model describes a naturally adaptive sensor: not only does cytoskeletal tension adjust according to the substrate stiffness, but this remodelling adapts the sensor response so that it remains most sensitive to its immediate surroundings -- small changes in the substrate stiffness will give large changes in the activation rate of FAK activation once the positive and negative feedback rebalance in homeostasis. 

This is desirable behaviour in a biological sensor, and it is remarkable that it is produced in our model with no prior stipulation. We initially only required that the cell be responsive to changes in the stiffness -- how big these changes were, or if they were optimal, was not close to the front of our minds. For such a simple model to predict useful adaptive sensing behaviour is exciting to us.

Our model of a single focal adhesion kinase is obviously not the whole story. There have been several experimental works showing that FAK dimerisation is an important initiator of FAK autophosphorylation. We did not attempt to capture any collective effects in the present model, and acknowledge that there is significant ground to be gained in expanding our model to a one describing the allosteric coupling. Nevertheless, one can easily see how such collective effects might be generated within our model.

Phosphorylated FAK acts on several important signalling molecules, such as Rho and Rac. If these molecules act to increase the tension in actin filaments in the broad vicinity, rather than strictly for filaments attached to active FAK molecules, then it is obvious that there will be a cooperative effect -- once a single focal adhesion kinase autophosphorylates, the tension in surrounding filaments will increase, and this increases the probability of a second opening event, and so on. 

The dependence of the FAK opening rate on stress relaxation partly explains results obtained in experimental work on cell spreading with different viscoelastic substrates~\cite{Chaudhuri}. Chaudhuri et al. saw suppression of cell spreading (associated with lower FAK activation) on substrates with significant stress relaxation, compared with purely elastic substrates of nominally the same storage modulus. We should note that we fail to capture the behaviour Chaudhuri et al. observed at very low stiffness (1.4kPa). On such a soft substrate, they saw that the number of cells with stress fibers was actually enhanced on substrates with stress relaxation -- the opposite trend to stiffer substrates. This isn't surprising; our model deals with mature focal adhesion complexes that would not develop on substrates of $\sim$kPa stiffness.

In summary, this work develops a theoretical model of the physical mechanism that a reversible mechanosensor of the 2nd kind should use. We focus all our discussion on the focal adhesion kinase, in association with integrin and talin, connecting the force-providing cytoskeletal F-actin and the varying-stiffness ECM. However, the fundamental principles of the model apply to all reversible molecular complexes that may be represented by the two-spring model of Fig. \ref{fig3new}. In this case, instead of FAK, it could instead (or simultaneously) be any of the other big molecules with autocatalytic activity involved in  the force chain in Fig. \ref{fig0}. The obvious alternative would be talin, which has the confirmed connection between integrin and actin \cite{Haller2016} and the large conformational change under applied force \cite{talin2016}. The next steps are to link the main result of this work (the rate of opening $K_+$) with the non-linear dynamics of one or several signalling pathways that produce the morphological response of the cell to the signal the mechanosensor generates.

\subsection*{Author contribution}
All authors contributed to carrying out research and writing the paper. 

\subsection*{Acknowledgments}
We have benefited from many useful discussions and support of G. Fraser, K. Chalut, T. Alliston, X. Hu, and D. C. W. Foo. This work has been funded by EPSRC EP/M508007/1.


\end{document}